\documentclass[prd,aps,
tightenlines,
superscriptaddress,showpacs,nofootinbib,eqsecnum,amsfonts,amsmath,epsf
]{revtex4}
\usepackage{bm,graphics,graphicx,epsfig,color
}

\def\gsim{\mathrel{
\rlap{\raise 0.511ex \hbox{$>$}}{\lower 0.511ex
\hbox{$\sim$}}}}
\def\lsim{\mathrel{
\rlap{\raise 0.511ex \hbox{$<$}}{\lower 0.511ex
\hbox{$\sim$}}}}

\def\i{\iota}

\def\ba{\begin{eqnarray}}
\def\ea{\end{eqnarray}}
\def\be{\begin{equation}}
\def\ee{\end{equation}}

\begin{document}
\hskip 2.75cm 
\title{ 
Higher harmonics increase LISA's mass reach for supermassive black holes
}

\author{K.G.\ Arun} \email{arun@lal.in2p3.fr} 
\affiliation{Raman Research Institute, Bangalore, 560 080, India}
\affiliation{LAL, Univ Paris-Sud, IN2P3/CNRS, Orsay, France}
\affiliation{${\mathcal{G}}{\mathbb{R}}
\varepsilon{\mathbb{C}}{\mathcal{O}}$, Institut d'Astrophysique de
Paris--- C.N.R.S.,
Paris, France} 
\author {Bala R Iyer} \email{bri@rri.res.in}
\affiliation{Raman Research Institute, Bangalore, 560 080, India}
\author{B S Sathyaprakash}
\email{B.S.Sathyaprakash@astro.cf.ac.uk}
\affiliation{School of Physics and Astronomy, Cardiff University, 5, The Parade, Cardiff, UK, CF24 3YB}
\author {Siddhartha Sinha$^{1,}$}\email {p_siddhartha@rri.res.in}
\affiliation{Dept. of Physics, Indian Institute of Science, Bangalore, 560 012, India.}
\begin{abstract}
Current expectations on the signal to noise ratios and masses of supermassive black holes which the
Laser Interferometer Space Antenna (LISA) can observe are based on using in 
matched filtering only
the dominant harmonic of the inspiral waveform at twice the orbital frequency.
Other harmonics will affect the signal-to-noise
ratio of systems currently believed to be observable by LISA. More
significantly, inclusion of other harmonics in our matched filters 
would mean that more massive
systems that were previously thought to be {\it not} visible in LISA should be detectable with reasonable SNRs. Our estimates show that we should be
able to significantly increase the mass reach of LISA and observe the more 
commonly occurring supermassive black holes of masses $\sim 10^8M_\odot.$
More specifically, with the inclusion of all known harmonics LISA will
be able to observe even supermassive black hole coalescences with total
mass $\sim 10^8 M_\odot (10^9M_\odot)$ (and mass-ratio  $0.1$)  
for a low frequency cut-off
of $10^{-4}{\rm Hz}$ $(10^{-5}{\rm Hz})$ with an SNR up to $\sim 60$ $(\sim 30)$  
at a distance of 3\ Gpc.  This is important from the
astrophysical viewpoint since observational evidence for the existence
of black holes in this mass range is quite strong and binaries
containing such supermassive black holes will be inaccessible to LISA  
if one uses as detection templates only the dominant harmonic. 
\end{abstract}
\date{\today}
\maketitle

\section{Introduction}\label{sec:intro}

\subsection{Supermassive black hole binaries and LISA}
There is strong observational evidence for the existence
of supermassive black holes (SMBHs) with masses in the 
range of $10^6M_\odot$--$10^9M_\odot$ (see e.g.\ Ref.\ \cite{FerrMerr00} and
references therein) in most galactic nuclei~\cite{Richstone98}. 
Therefore, mergers of galaxies, as evidenced by high-redshift surveys, should give
rise to  binaries containing  SMBHs. Late stage evolution of a SMBH binary 
is dictated by the emission of gravitational radiation. The resulting loss of
orbital energy and angular momentum would lead to the coalescence of the two holes.
Indeed, X-ray observations have revealed the existence of at least one such system 
that would coalesce within the Hubble time \cite{KomossaEtAl}.  Gravitational waves (GW) emitted
in the process could be detected by the planned 
Laser Interferometer Space Antenna (LISA) \cite{lisa}. 

Observation of SMBH binaries at high redshifts is one of the
major science goals of LISA. These observations will
allow us to probe the evolution of SMBHs and structure
formation~\cite{Hughes02} and provide an unique opportunity 
to test General Relativity (and its alternatives) in the strong 
field regime of the theory~\cite{BBW05a,BBW05b,AIQS06a,AIQS06b,HughMen05}.
Observing SMBH coalescences with high ($100$-$1000$) SNR \cite{AIQS06a,AIQS06b} 
is crucial for performing all the aforementioned tests.

\subsection{Restricted Vs Full Waveforms as Search templates in LISA}

Motivated by the fact that matched filtering is more sensitive to the 
phase of the signal than its amplitude \cite{Cutler:1992tc}, 
search algorithms so far have deployed 
a waveform model involving only the  {\em dominant harmonic} (at twice the 
orbital frequency), although the phase evolution itself is included 
to the maximum available post-Newtonian (PN) order (currently 3.5PN, 
for non-spinning systems \cite{BFIJ02, BDEI04}).  Waveforms
in which all {\em amplitude} corrections are neglected, but the {\em phase} is
treated to the maximum available order, are
called {\em restricted waveforms} (RWF) and these are what are
used so far in the analysis of data from ground-based detectors
\cite{3mn,CF94,BDIWW95,B96}. This paper will consider the advantage
of using the {\em full wave forms} (FWF) in the context of LISA.
 
LISA is designed to detect gravitational waves in the frequency-band 
$0.1$--$100\, \rm mHz$. This frequency range determines the range
of masses accessible to LISA because the inspiral signal would end
when the system's orbital frequency reaches the mass-dependent
last stable orbit (LSO).
In the test-mass approximation, the  angular velocity $\omega_{\rm LSO}$
at LSO is given by $\omega_{\rm LSO} = 6^{-3/2}M^{-1},$ 
where $M$ is the total mass of the binary. Search templates that contain
only the dominant harmonic cannot extract power in the signal beyond $f_{\rm LSO} = 
\omega_{\rm LSO}/ \pi \simeq 4.39 (M/10^6 M_\odot)^{-1} \rm mHz.$ 
This further implies that the frequency range $[0.1,\, 100]\,\rm mHz$ 
corresponds to the range
$\sim 4.39\times [10^4,\,10^7]M_\odot$ for the total mass of 
binary black holes that would be accessible to LISA\footnote{Although, 
binaries lighter than $10^4M_\odot$
would, in principle, evolve through the LISA band they would not
be luminous enough to be visible in LISA unless they are close-by.}.
However, as Table 1 of Ref.~\cite{FerrMerr00} would reveal, 
there is observational evidence for the existence of many SMBHs 
whose masses are of the order of $10^8$--$10^9M_\odot$.
LISA will be {\it unable} to observe binaries containing SMBHs in this 
mass range if it used as search templates waveforms containing {\it only}
the dominant harmonic.

Inclusion of  higher-order amplitude terms in the
waveform introduces the following two new features: 
(i)~appearance of higher harmonics of the orbital phase 
and (ii)~PN amplitude corrections to the leading
as well as higher harmonics of the orbital frequency.
For example, at 0.5PN order, which is the first-order correction, there are
two new harmonics $\Psi$ and $3\Psi$, where $\Psi$ is related to the orbital
phase of the binary as in Refs.~\cite{BIWW96,ABIQ04}. 
More interestingly, in the expressions for the `plus' and `cross' polarizations,
all odd harmonics of the orbital frequency are
proportional to $\frac{\delta m}{M}$, where $\delta m$ is the difference
in the masses of the binary components
(see Eq.~(5.7)-(5.10) of Refs.~\cite{ABIQ04}).
Another important feature of the  full waveform is that 
the $(2 n+2)^{\rm th}$ harmonic first appears 
at the $n^{\rm th}$ PN order in amplitude\footnote{The 0.5PN term is an exception to this
and also introduces a harmonic at the orbital frequency apart
from the one at thrice the orbital frequency.}.  
For example, the fourth harmonic first appears at $1$PN, and has
PN amplitude corrections to its dominant term at $2$PN and $2.5$PN 
(see Refs.~\cite{ABIQ04, BIWW96} for details).

Early investigations on the importance of amplitude-corrections to 
search templates were carried out by Sintes and 
Vecchio~\cite{SinVecc00a,SinVecc00b}. Their study used only
the first-order correction at $0.5$ PN order. They concluded that 
the addition of the amplitude terms in the waveform
did not improve the accuracy in the estimation of source's
angular position and the distance, whereas the estimation of 
the chirp and reduced masses could be 10 times better when compared 
to the RWF. Recently, in the context of ground-based detectors, 
Van Den Broeck and Sengupta~\cite{Chris06,ChrisAnand06,ChrisAnand06b} 
examined the implications of going beyond the restricted PN approximation 
and employing instead the full waveform  \cite{BIWW96,ABIQ04}.  
The two main implications of  the comprehensive
analysis in  Refs.~\cite{Chris06,ChrisAnand06}
for terrestrial GW detectors may be summarized as follows:
\begin{enumerate}
\item 
For binary neutron stars and stellar mass black holes, 
restricted waveforms over-estimate the SNR as compared 
the full waveform.
\item The use of the full waveforms significantly increases the mass-reach of 
second and third generation detectors, advanced LIGO and EGO
being able to observe systems with total mass $\sim 400 M_\odot$ and
a third generation detector as high as $10^3 M_\odot.$
\end{enumerate}
     
In the present paper, we study in the context of LISA
the implication of using
templates based on the  FWF (i.e. including
{\it all} known harmonics of the orbital phase and {\it all} 
known amplitude corrections in the GW polarisations).
Coalescences of SMBH binaries with masses 
$\sim 10^{8-9}M_\odot$ will {\it not} be observable by LISA
if one uses only templates based on the RWF. 
Using  templates based on amplitude corrected full waveforms,
instead of the usual restricted waveforms,
will enable LISA to observe coalescences of SMBH binaries
with total mass $\sim 10^8 M_\odot$ $(10^9M_\odot)$ if the lower
frequency cut-off LISA can achieve is $\sim 10^{-4}$Hz ($10^{-5}$Hz).  

The rest of this paper is organized as  follows:
In Section \ref{FWF}, we give the FWF in the frequency domain, 
by taking into account 
the orbital motion of LISA around the sun and its changing orientation.
Section \ref{LISA-SNR} discusses the results of our investigations 
where we compare the performances of  the amplitude-corrected waveforms 
at different PN orders in terms of their mass-reach and 
distance-reach and correlate it to the `observed' spectrum in LISA. 
Section \ref{Summary} concludes with a brief summary 
of the main results and assumptions underlying their derivation.

\section{Template waveforms for LISA}
\label{FWF}

\subsection{Amplitude corrected waveform}

For non-spinning binaries in quasi-circular orbits inspiralling 
due to radiation-reaction, waveforms were computed in 
Refs.\ \cite{BIWW96,ABIQ04} up to 2.5PN order in amplitude and 3.5PN 
in phase~\cite{BFIJ02,BDEI04}. This waveform $h(t)$ 
is a linear combination of sine and cosine functions of  
multiples of the orbital phase $\Psi(t)$.  The expression 
for the 2.5PN polarization contains the first seven harmonics 
of the orbital phase, the dominant harmonic being the one at 
twice the orbital phase.  The signal depends on the following 
parameters: $D_L$, the luminosity distance to the binary, $m$ 
the total (red-shifted) mass, $\nu$ the symmetric mass-ratio 
(reduced mass divided by total mass), the spherical polar angles  
$(\theta,\phi)$ determining the direction of the ``line-of-sight'', 
the inclination angle $\iota$ of the angular momentum $\bf{ L} $ 
of the binary with respect to the direction opposite to the line-of-sight, 
and the polarization angle $\psi$ which determines the orientation of 
the projection of  $\bf{ L} $ in the plane normal to the 
line-of-sight.
 
We rewrite the waveform in terms of only  cosines in a form similar to
 \cite{ChrisAnand06}:
\begin{equation}
h(t) = \frac{2M\nu}{D_L}  \,\sum_{k=1}^{7} \sum_{n=0}^{5}
A_{(k,n/2)} \cos[k\Psi(t) +\phi_{(k,n/2)}]\,x^{\frac{n}{2}+1}(t),\label{ht}
\end{equation}
where the coefficients $A_{(k,n/2)}$ and $ \phi_{(k,n/2)}$  are 
functions of $(\nu,\theta,\phi,\psi,\i)$, and  
$x(t)=(2\pi M F(t))^{2/3}$ is the post-Newtonian parameter 
with $F(t)$ the instantaneous {\it orbital} frequency. 
Terms $\frac{2M\nu}{D_L} x^{n/2+1}(t)\, A_{(k,n/2)}$  and $\phi_{(k,n/2)}$
are the wave amplitude and polarization phase, respectively, corresponding
to the  $k^{\rm th}$ harmonic and $(n/2)^{\rm th}$ PN order. We call the
coefficients $A_{(k,n/2)}$ the polarization amplitudes.
The orbital phase $\Psi(t)$ is a PN  series in $x$, which, in 
the case of non-spinning binaries, is known to 3.5PN order~\cite{BFIJ02, BDEI04}. 
For a non-spinning source and a detector whose position and 
orientation are almost constant during the time of observation 
of the signal, all the above mentioned angles are constants. 
For ground-based GW detectors dealt with in 
Ref.~\cite{Chris06,ChrisAnand06}, one is in this situation.
\subsection{Amplitude corrected waveform
including modulations due to LISA's orbital motion -- Time Domain }
\label{LISA-FWF-TD}
LISA will be able to observe many sources from their early stages
of inspiral and most would last for a pretty long time. We shall only 
consider binary sources that last for a year or less before merger.
Since the LISA plane is tilted by $60^\circ$ with respect to the 
plane of the ecliptic, during the course of its heliocentric orbit its
orientation and position varies periodically, with a period of one 
year and the signal in Eq.~(\ref{ht}) will suffer
additional amplitude and phase modulations. 
Thus in the case of LISA  the angles $\theta$, $\phi$, and $\psi$
(but not $\i$) appearing in Eq.~\ref{ht} are functions of time.
 To proceed further,
in the frame of a non-rotating observer fixed to the solar-system
  barycentre,
we denote by the location of the source on the sky by the
spherical polar angles $\theta_S$ 
and $\phi_S$ 
and the orientation of the source 
by the spherical polar angles $\theta_L$ and $\phi_L$ determining
the direction of the orbital angular momentum $\bf{L}$ of the binary.
 The transformation between the fixed set of 
angles\footnote {This is a different notation from \cite{Cutler98}, where the
source angles measured in the fixed  barycentre frame are denoted by 
($\bar \theta_{\rm S},\, \bar \phi_{\rm S},\, 
\bar \theta_{\rm L},\, \bar \phi_{\rm L} $)}
($\theta_S,\,\phi_S,\,\theta_L,\,\phi_L$) and the time-dependent angular 
coordinates of the source $(\theta(t),\,\phi(t),\psi(t),\i)$ 
as measured by LISA are  given in Ref.\ \cite{Cutler98}. 

Generalizing, 
Ref.\ \cite{Cutler98} from the RWF to the FWF,
 the signal as seen in LISA is of the form,
\begin{equation}
h(t) =\frac{\sqrt{3}}{2}\, \frac{2M\nu}{D_L}  \,
\sum_{k=1}^{7} \sum_{n=0}^{5}\, A_{(k,n/2)}(t) 
\cos[k\Psi(t)+\phi_{(k,n/2)}(t)+k\phi_{D}(t)] \,x^{\frac{n}{2}+1}(t)\,.
\label{eq:Doppler wf}
\end{equation}
The PN parameter $x(t)$ appearing in Eq.~(\ref{eq:Doppler wf}) is still
equal to $(2\pi M F(t))^{2/3}$, where $F(t),$ however, is the orbital frequency 
as measured by a {\it non-rotating observer located at the solar-system barycentre}.
The term $\phi_{D}(t)$ is the {\it Doppler phase}~\cite{Cutler98}, accounting
for the  phase difference of the gravitational wave-front between 
LISA and the solar-system barycentre. The time-dependence of $\phi_{D}(t)$ 
is due to the orbital motion of LISA about the barycentre. It is given by
\begin{equation}
\phi_{D}(t)=2\,\pi\,F(t)\,R\,\sin\theta_{S}\cos[\phi(t)-\phi_{S}],\label{phiD}
\end{equation} 
where $R = 1$ AU    
and $\phi(t)$ is the angular position of LISA with 
respect to the barycentre given by $\phi(t)= 2\,\pi\,\frac{t}{T},$ $T$ 
being equal to one year.
\subsection{Amplitude corrected waveform
including modulations due to LISA's orbit -- Frequency Domain }
\label{LISA-FWF-FD}
The above waveform is valid in the adiabatic regime, where the radiation-reaction 
time-scale is much larger than the orbital time-scale.  We also note that 
the additional amplitude and Doppler modulations in the waveform for LISA 
vary on time-scales of 1 yr (i.e. $\sim 3 \times 10^7\,\rm s$), while LISA
can observe orbital periods at most up to $2 \times 10^{5}\,\rm s,$ (i.e. gravitational wave frequencies of order $10^{-5}\, \rm Hz.$).  
Consequently, the Doppler modulations change much more slowly 
(a hundredth) than the orbital phase.
This permits the use of the stationary phase approximation (SPA) to obtain 
an analytical form for the Fourier transform (FT) $\tilde{h}(f)$ of the signal:
\begin {equation}
\tilde{h}(f) \simeq \frac{\sqrt{3}}{2} \frac{2M\nu}{D_L} \,\sum_{k=1}^{7}\,\sum_{n=0}^5\,
\frac{A_{(k,n/2)}(t(f/k)) \,x^{\frac{n}{2}+1}(t(f/k))\,e^{-i\phi_{(k,n/2)}(t(f/k))}}{2\sqrt{k\dot{F}(t(f/k))}}\,
\exp\left[i\,\psi_{ f}(t(f/k))\right],
\label {FT}
\end {equation}
where an over dot denotes derivative with respect to time and  
$\psi_f(t(f/k))$ is given by
\begin {equation}
\psi_f(t(f/k)) = 2\pi f\,t(f/k) -
k\,\Psi(t(f/k))-k\,\phi_{D}(t(f/k)) - \pi/4.
\label {phase}
\end {equation}
The PN expansions for $t(F),\Psi(F),\dot{F}(F)$ are
given in Ref.\ \cite{BFIJ02}. The expression in Eq.~(\ref{FT}) within the summation
over $k$ represents the FT due to the $k^{\rm th}$ harmonic. 
It should be noted that the term $\dot{F}$ may be treated in different
ways that could lead to numerically different results. In a numerical 
treatment, for instance, one could avoid performing a further
re-expansion.  Alternatively, one could re-expand the denominator
in the amplitude  and truncate the resulting expression at the
$n^{\rm th}$ PN order, to obtain the $n$PN amplitude-corrected waveform.
Ref.~\cite{Chris06,ChrisAnand06} choose the latter and
we follow them in this work.

Radiation reaction results in an increase in the orbital frequency $F(t)$ 
which will ultimately drive the system beyond the adiabatic inspiral 
phase and the inspiral waveform given above will no longer be valid.  
In the first approximation this is expected to occur
 when the orbital frequency 
$F(t)$ reaches $F_{\rm LSO}$ -- the orbital frequency of
the LSO of a Schwarzschild solution with the same 
mass as the binary's total mass $M$,
\begin{equation}
F_{\rm LSO}=(2\,\pi\,6^{\frac{3}{2}}\,M)^{-1}.
\end{equation}
Thus, we truncate the signal
in the time domain  at  a time $t_{\rm LSO}$, given implicitly by
$F(t_{\rm LSO})= F_{\rm LSO}$.
In the SPA, the main contribution to
the FT of  the $k^{\rm th}$ harmonic at a given Fourier frequency $f$,
comes from  the  neighbourhood of the time when the instantaneous value of the
$k^{\rm th}$ harmonic sweeps past $f$. 
Thus the $k^{\rm th}$ harmonic is not expected to
contribute significant power to the FT for frequencies above $k\,F_{\rm
LSO}$, if the signal is truncated in the time domain beyond  $t_{\rm LSO}$.
This  motivates the truncation of the FT due to the $k^{\rm th}$ harmonic
at frequencies above  $k\,F_{\rm LSO}$ by a step function $\theta(k\,F_{\rm LSO}-f)$
[$\theta(x)=1,$ for $x\ge 0$, and $0$ for negative $x$].

\section{Signal to noise ratios in LISA with higher harmonics}
\label{LISA-SNR}
In this Section we investigate the effect of the higher harmonics
in LISA observations of supermassive black hole binaries. 
The LISA waveform discussed in the previous Section will be used
for the analysis.

Given a waveform $h$, the best signal-to-noise ratio (SNR) achieved using
an optimal filter is given by $\rho[h] \equiv (h|h)^{1/2}$, 
where $(\,.\,|\,.\,)$ is the usual inner product in terms of the 
one-sided noise power spectral density $S_h(f)$ of the 
detector. With the convention for Fourier transforms,
$\tilde{x}(f) = \int_{-\infty}^{\infty} x(t)\,\exp(-2\pi i f t)\,dt$,
the inner product is given by: 
\begin{equation}
(x|y)  \equiv 4\int_{f_{\rm s}}^{f_{\rm end}} \frac{\mbox{Re}[\tilde{x}^\ast (f)\tilde{y}(f)]}{S_h(f)}df.
\label{innerproduct}
\end{equation}
For an optimal filter, which maximises the overlap of the
signal with template, one can write
\begin{equation}
\rho^2 = 4\int_{f_{\rm s}}^{f_{\rm end}} \frac{|\tilde{h} (f)|^2}{S_h(f)}df.
\label{snr}
\end{equation}
We use the non-sky-averaged noise-spectral-density as given 
in Eqs.~(2.28)-(2.32) of Ref.~\cite{BBW05a}. 
\subsection{Choice of frequency cutoffs $\mathbf{f_{\rm end}, f_{\rm s}}$}
The upper limit of integration $f_{\rm end}$ 
is taken to be the 
minimum of $7\,F_{\rm LSO}$ and $1$ Hz, the latter being a conventional upper 
cut-off for the LISA noise curve. The lower limit $f_{\rm s}$ is chosen assuming 
LISA observes the inspiral for a duration $\Delta t_{\rm obs}$
before it reaches the LSO.
Using the quadrupole formula, we find that the orbital frequency at the epoch
$t_{\rm LSO} -\Delta t_{\rm obs}$ is given by 
\begin{equation}
F(t_{\rm LSO}-\Delta t_{\rm obs})= \frac{F_{\rm LSO}}{(1+\frac{256\, \nu}{5\,M}\,
\label{eq:Fin}
\Delta t_{\rm obs}\,v_{\rm LSO}^8)^{\frac{3}{8}}},
\end{equation}
where $v_{\rm LSO}$ is the orbital velocity and $t_{\rm LSO}$ 
the epoch at which the orbital frequency reaches the value
$F_{\rm LSO}$. We take $v_{\rm LSO}=1/\sqrt{6},$ 
the orbital velocity at the LSO in the case of a test mass
orbiting a Schwarzschild black hole. We designate 
$F(t_{\rm LSO}-\Delta t_{\rm obs})$ as $F_{\rm in}.$ 
Thus the $k^{\rm th}$ harmonic will have a 
frequency $k\,F_{\rm in}$, $\Delta t_{\rm obs}$  before $t_{\rm LSO}$. 
The above formula reduces to the simpler form
in Ref.\ \cite{BBW05a} as $v_{\rm LSO} \rightarrow \infty$.
For the mass values explored in this work there is no significant
dependence of the results on this choice. In all our calculations we take $\Delta t_{\rm obs}$ to be one year.

The lower cut-off for the $k^{\rm th}$ harmonic
should be the maximum of the lower cut-off
of LISA ($10^{-4}$ Hz) and $k\, F_{\rm in}$
and simply implemented
by truncating the waveform due to the $k$th harmonic
by  another step-function  $\theta(f-k\,F_{\rm in})$
and choosing $f_{\rm s}$  to be $10^{-4}$Hz.
It is worth noting that  the $k^{\rm th}$ harmonic probes a 
larger interval  of the frequency
domain i.e.   $k(F_{\rm LSO}-F_{\rm in})$ relative to the fundamental
harmonic. For brevity, we refer to this as the span of the 
$k^{\rm th}$ harmonic.
  
There is a caveat with regard to the use of higher harmonics
that is worth mentioning: In the time-domain the waveform
should begin when the highest harmonic reaches the 
lower cutoff.  This has an implication on data analysis as the 
templates will be an order-of-magnitude longer than before. 
Thus, it might be sensible to use higher harmonics only in the
case of higher masses.

\begin{figure*}[t]
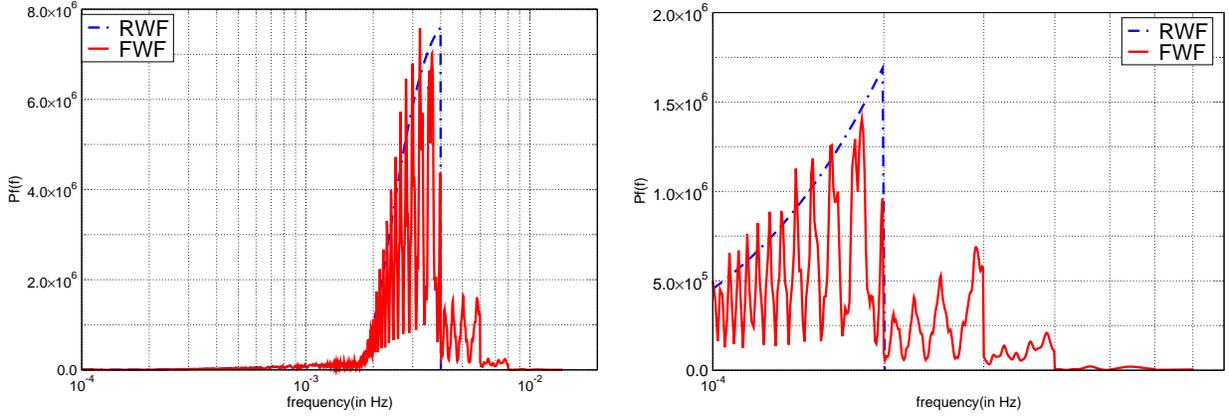

\centering
\includegraphics[width=3.1 in]{Pf5_6.eps}
\hskip 0.35 true cm
\includegraphics[width=3.1 in]{Pf2_6_7.eps}
\caption{The observed spectrum, ${\cal P}(f)\equiv \frac{d(\rho^2)}{d(\ln f)}= \frac{f\,|\tilde{h}(f)|^2}{S_h(f)}$, in LISA using the full (solid, red) and
restricted (dashed, blue) waveforms, for two archetypal binaries,
$(10^5,\, 10^6) M_\odot$ (left) and $2 \times (10^6,\, 10^7) M_\odot$ (right). 
The sources are assumed to be at 3 Gpc and their orientation 
with respect to the solar-system barycentre is chosen to be 
$\theta_S=\cos^{-1}(-0.6)$, $\phi_S=1$, $\theta_L=\cos^{-1}(0.2)$, $\phi_L=3$.
The spectrum is much more complicated and highly oscillatory
for the FWF than for the RWF, because of interference 
between various harmonics. The higher frequency
reach of the FWF is due to presence of higher harmonics as
apparent in the figure. 
The spectrum for the system in the left panel sharply rises 
at a frequency $\sim 2 \times 10^{-3}$Hz. 
Beyond this frequency, the effective LISA noise decreases sharply 
with increasing frequency (as there are fewer galactic binaries per 
frequency bin) leading to the observed increase in the spectrum.
}
\label{spectrum}
\end{figure*}

\subsection{Observed signal spectrum with LISA}
To get some insight into the effect of higher harmonics via amplitude 
corrections let us first look at the SNR integrand,  {\it i.e.}, the 
``noise-weighted signal power'' per unit logarithmic frequency 
interval \cite{DIS00}. Rewriting the expression for the SNR as
\begin{eqnarray}
\rho^2 
       &=& 4\int_{f_s}^{f_{\rm end}} \frac{f\,|\tilde{h} (f)|^2}{S_h(f)}d\,{\rm ln}(f),
\end{eqnarray}
the quantity of our interest is
\begin{equation} 
{\cal P}(f)\equiv \frac{d(\rho^2)}{d(\ln f)}= \frac{f\,|\tilde{h}(f)|^2}{S_h(f)}, 
\end{equation} 
which is designated as the
``observed spectrum", following \cite {ChrisAnand06b}.
The observed spectrum is plotted versus frequency for given
masses in Fig.~\ref{spectrum}. As is the case for ground-based detectors
\cite {ChrisAnand06b}, the spectrum due to the FWF has a lot more structure
and is highly oscillatory because of interference  between various
harmonics. For the $(10^5,\, 10^6) M_\odot$ system, the mass being low,
 the second harmonic and hence the RWF extends up to frequencies
$\sim 2 \times 10^{-3}$ Hz, where LISA is most sensitive. This leads to a
rapid increase in the observed spectrum  in this frequency region. The spectrum
due to the FWF, containing higher harmonics continue beyond the RWF 
into the most sensitive part of the
LISA band. For the $2(10^6,\, 10^7) M_\odot$ system, the frequency span of the 
second harmonic is small and the sensitive region of the LISA band lies beyond 
 its maximum reach.

\begin{table}[t]
\centering
\caption{SNRs due to successive PN amplitude-corrected waveforms, 
with phase corrections to  $3.5$ PN order in all cases. 
The orientation of the source with respect to the solar-system barycentre is
chosen to be  $\theta_S=\cos^{-1}(-0.6)$, $\phi_S=1$, $\theta_L=\cos^{-1}(0.2)$, $\phi_L=3$.
 For the $(10^6-10^7)M_\odot$  binary system, 
all harmonics enter deep into the sensitive part of the LISA bandwidth.
Apart from an increase at 0.5PN, 
we see a consistent reduction in the SNR on inclusion of higher PN order 
amplitude corrections. 
For the $(5.5\times10^6 ,5.5\times10^7) M_{\odot}$ binary system,
 the second  harmonic fails to enter the LISA bandwidth, 
while the third harmonic spans a small insensitive region. 
Thus the SNR due to the RWF is zero, while the SNR due to the $0.5$PN waveform
is smaller than the SNRs due to higher order PN terms. 
Both sources are at a  distance of 3 Gpc.}
\label{snrtable}
\vskip 12pt
\begin{tabular}{|l|c|c|}
\hline
\hline
 PN  & \multicolumn {2}{c|}{SNR}\\
\cline{2-3}
order& \multicolumn{1}{c|}{~~~~~~($10^6-10^7) M_\odot$} &
\multicolumn{1}{c|}{$5.5\times(10^6-10^7)M_\odot$}\\
\hline
0   & 924.48  & 0      \\
0.5 & 1025.8  & 211.98 \\
1   & 928.48  & 343.17 \\
1.5 & 869.78  & 319.34 \\
2   & 824.65  & 266.65 \\
2.5 & 809.51  & 277.34 \\
\hline
\hline
\end{tabular}
\end{table}

\section{The effect of higher harmonics}

Following the analysis of
Ref.~\cite{Chris06,ChrisAnand06}, we classify the sources
into two types: In the first category are sources for which the 
dominant (second) harmonic has a large frequency span in the  
LISA band.  The second category on the other hand comprise 
sources whose dominant harmonic {\it{ fails}} to enter the LISA bandwidth
but the  higher harmonics {\it{do}}. Since the upper cut-off frequency 
for each harmonic is inversely proportional to the total mass 
(from the expression for $F_{\rm LSO})$, we note that the sources 
of the first type will have total mass less than some  value
which we call the RWF mass-reach, the maximum mass detectable by
the RWF, while the second type will have masses greater than 
this value. The condition that the upper cut-off of the dominant harmonic is 
less than or equal to the lower cut-off of LISA (i.e., by the 
inequality $2\,F_{\rm LSO}\le f_{\rm s}$) determines the RWF mass-reach.
The choice of $f_{\rm s}$ for the LISA mission is still not clear and theoretical
implications of this choice are explored in e.g.\ Ref.\ \cite {BC05}.
For $f_{\rm s}$ in the range $[10^{-5}\,, 10^{-4}]$ Hz 
the RWF mass-reach varies over the range $[4.39\,,43.9]\times 10^{7}M_\odot$,
the lower end of the mass range corresponding to the higher end of the
frequency range.
          
\subsection{How higher harmonics affect signal visibility}

\begin{figure*}[t]
\centering
\includegraphics[width=3.1in]{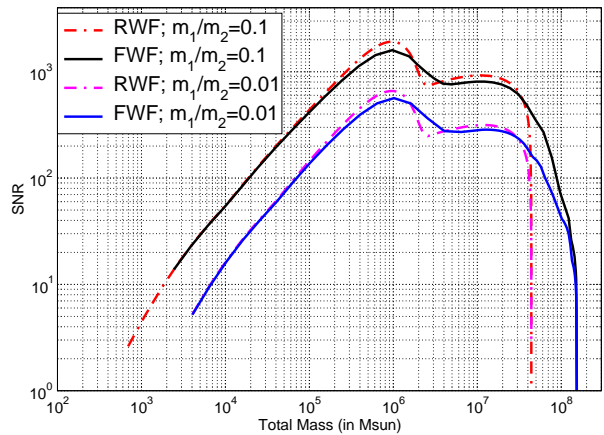}
\hskip 0.35 true cm
\caption{ SNR versus total mass for mass ratios of 0.1 and 0.01. 
The figure shows that apart from the dips due to white-dwarf confusion noise, 
for mass values where the RWF enters the  LISA band, the corresponding
SNR  is consistently more than the SNR produced by the FWF. 
However, for mass values where  the second harmonic terminates
 before it reaches the LISA bandwidth, the FWF 
which has higher harmonics that  enter the LISA band produces
 significant SNRs. 
The frequency reach of a harmonic  depends only on the total mass, 
and thus the mass reach of the FWF is independent of  the mass-ratio. 
For more asymmetric systems, 
the magnitude of the SNR is low for all masses both for the RWF and the FWF.
Sources are at a luminosity distance of 3 Gpc with fixed angles given by 
$\theta_S=\cos^{-1}(-0.6)$, $\phi_S =1$, $\theta_L=\cos^{-1}(0.2)$, $\phi_L=3$.}
\label{SNR-mass}
\end{figure*}

In Fig.\ \ref{SNR-mass} we plot the SNRs computed using 
the {\em restricted} (RWF) and {\em full} (FWF) waveforms as a 
function of the binary's total mass for two values of the mass 
ratio\footnote{Our codes are calibrated by reproducing 
the results of \cite{Chris06,ChrisAnand06}, which considers 
ground-based detectors, and also of \cite{ALISA06}, which 
computes SNRs in LISA using RWF.}.  We first consider  systems
whose total mass is less than $ 4\times 10^7 M_\odot$. For these
systems, the SNRs computed using the two different approximations 
agree with each other to within 10\%, with the RWF over-estimating 
the SNR, when compared to the FWF, in most of the range. This is 
explicitly shown for a $(10^6,\, 10^7)M_\odot$ binary in the first column 
of Table \ref{snrtable}.  Indeed, but for the slight increase 
in SNR as we go from 0PN to 0.5PN, we find a steady decrease  as 
one increases the PN order of the amplitude correction.  

The reduction in SNR at higher PN orders can be understood by 
studying the structure of $|\tilde{h} (f)|^2 $, the numerator 
in the integrand of the SNR in Eq.~(\ref{snr}). There are basically
three types of terms:
\begin{enumerate}
\item {\em direct} terms in which the phases in Eq.~(\ref{FT}) cancel 
$$ A_{(k,n/2)}^2(t(f/k))\,f^{-\frac{7}{3}}\,(Mf)^{\frac{2n}{3}},$$

\item {\em interference} terms between {\it different} PN corrections of the 
{\it same} harmonic,
$$A_{(k,m/2)}(t(f/k))\,A_{(k,n/2)}(t(f/k))\,f^{-\frac{7}{3}}\,(Mf)^{\frac{m+n
}{3}}\,\cos[\phi_{(k,m/2)}(t(f/k))-\phi_{(k,n/2)}(t(f/k))]$$

\item {\em harmonic mixtures\footnote{We use the term 
`harmonic mixtures' at the risk of being mistaken to the well-known 
`harmonic mixing' in music. Our use of the phrase `harmonic mixtures' 
is simply to convey the physical effect of the interference between different
harmonics}} which are terms containing the interference 
between {\it different} PN corrections of {\it different} harmonics, e.g. 
the $m/2^{\rm th}$ PN correction of  the $k^{\rm th}$ harmonic and 
$n/2^{\rm th}$ PN correction of the $l^{\rm th}$ harmonic. 
$$A_{(k,m/2)}(t(f/k))\,A_{(l,n/2)}(t(f/l))\,f^{-\frac{7}{3}}\,(Mf)^{\frac{m+n}
{3}}\, \cos[\psi_f(t(f/k))-\phi_{(k,m/2)}(t(f/k))-
\psi_f(t(f/l))+\phi_{(l,n/2)}(t(f/l))]$$
where $\psi_f(t(f/k))$ is given by Eq.~(\ref{phase}),
\end{enumerate}
All these terms are scaled by ${\cal M}^{5/3}$,
where ${\cal M}=M\,\nu^{3/5}$ is the  chirp-mass.
(Additional multiplicative factors have been omitted in the above expressions,
among which are the step-functions mentioned earlier and PN
expansion coefficients of the denominator of the Fourier amplitude in
Eq.~(\ref{FT}), the latter being time-independent.)

\subsection{The effect of higher harmonics in ground-based detectors}
Before we explain the SNR trends in the context of LISA, we mention that for
ground-based detectors a similar effect was found in Ref.\ \cite{ChrisAnand06} 
for a different but corresponding mass region. 
The lower cut-off for a typical ground-based
detector, say Advanced LIGO is $20$Hz, and the effect of higher harmonics is 
seen for masses less than $\sim 220 M_{\odot}$.
In that case, as mentioned earlier, the polarisation amplitudes and phases
are constants. The RWF contains only the Newtonian term of the second
harmonic and thus $|\tilde{h} (f)|^2 $ consists of a single direct term with
$n=0$ and $k=2$. 

With the inclusion of higher-order amplitude terms in the waveform, PN
corrections to the dominant harmonic, and higher harmonics and their PN
corrections, also contribute to the SNR.  In other words, the signal power 
spectrum $|\tilde{h} (f)|^2 $  will contain all three types of terms discussed 
before. From the form of the {\em direct terms,} it is evident that their
contribution to the SNR will be positive definite. We also note that, for
ground-based detectors, the frequency dependence of the {\em direct \rm and 
\em interference} terms will  just be a power law. However, the sign of the 
interference terms (and consequently their contribution to the SNR) depends on the
difference between the polarisation phases of different PN corrections for
the same harmonic. Van Den Broeck and Sengupta showed that for a given
harmonic, for all allowed values of the parameters $(\nu,\theta,\phi,\psi,\i)$, 
each PN correction is almost ``out of phase'' with {\em both} the PN correction 
preceding and succeeding it\footnote{Note, however, that Ref.\ \cite{ChrisAnand06}, 
argues this in a somewhat different form.}.  The resulting negative terms (representing 
destructive interferences) reduce the SNR as one includes higher PN amplitude
corrections in the waveform. 

The third type of terms, harmonic mixtures, however, are 
highly oscillatory functions of the frequency, as the phase difference 
$\psi_f(t(f/k))-\psi_f(t(f/l))$ between the $k^{\rm th}$ and the $l^{\rm th}$ 
harmonic become even or odd multiples of $\pi$.  As one integrates over $f$,
these oscillations tend to cancel out, and thus the contribution to the SNR
from these terms are numerically much smaller relative to the first two 
types of terms. 

\begin{figure*}[t]
\centering
\includegraphics[width=3.1in]{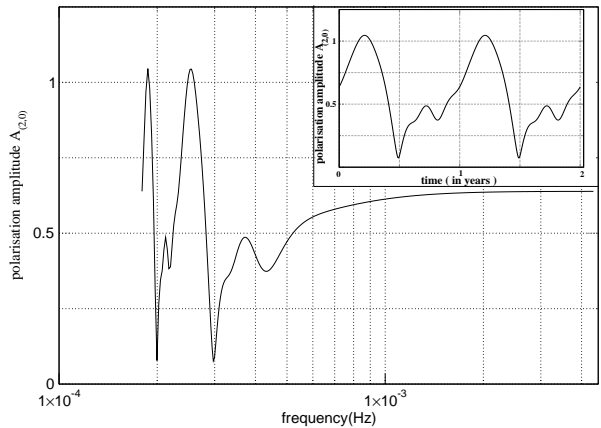} 
\hskip 0.35 true cm
\caption {Variation of polarisation amplitude of the RWF with frequency and 
time (inset). The inset, plotted over a duration of two years clearly
shows periodicity due to LISA's orbital motion around the Sun. The binary
mass, ($10^6-10^4)M_{\odot}$, has been chosen such that it can, in principle,
be observed for two years. The plot in the frequency domain shows that the 
variation of the polarisation amplitude is confined to a very small part
of the frequency span of the dominant harmonic, and essentially behaves
as a constant in the frequency domain.} 
\label{freq-time}
\end{figure*}  
\subsection{Effect of higher harmonics for binaries with $\mathbf{M<4 \times 10^7M_\odot}$}

In the case of LISA, because of the polarisation factors, 
the amplitudes of none of the three types of terms is a simple power-law in $f$.
The periodic variation of, for example, $A_{(2,0)}$ (period being one year)
appears as an amplitude modulation $A_{(2,0)}(t(f/2))$ in the Fourier transform, 
where the argument $t(f/2)$ of $A_{2,0}$ is given by
\begin{equation}
t(f/2)=-\frac{5}{256\pi^{8/3}{\cal M}^{5/3}}\frac{1}{f^{8/3}}+ \rm {PN\,
corrections}.
\label{eq:toff}
\end{equation}
Hence, in the frequency domain $A_{(2,0)}$ will undergo one complete
oscillation as  $f$ varies from $2F_{\rm in}$ (see Eq.~(\ref{eq:Fin})) to 
$2F_{\rm LSO}$. However, because of the {\it{inverse}} power-law dependence on $f$,
the oscillation of  $A_{(2,0)}$ is confined to a small frequency interval
above  $F_{\rm in}$ and remains fairly constant over a major portion of the 
frequency span $2(F_{\rm LSO}-F_{\rm in})$ (see Fig.\ \ref{freq-time}).
For masses higher than the one shown in Fig.\ \ref{freq-time},
this region of significant variation moves to the left of the
figure. On including in our analysis the effect of 
detector sensitivity (weighting down by $S_h(f)$) this variation of
$A_{(2,0)}$ gets damped out when one evaluates the integral in
Eq.\ (\ref{snr}).  For masses satisfying  $2F_{\rm in}\ll 10^{-4}$ Hz,
the lower cut-off for LISA, this region of variation will fall
below the LISA band. 

The polarisation phases determining the sign of the interference
terms between the same harmonics also vary with $f$. However, as mentioned
earlier, the phase relationships of the polarisation phases are independent
of the parameter values. Thus the modulations which change the values
of $(\theta,\phi,\psi,\i)$ do not affect the trend of reduction of SNR with
amplitude corrections. The Doppler modulations, which appear in only 
harmonic mixtures, are also not important as far as SNR is concerned.

\begin{figure*}[t]
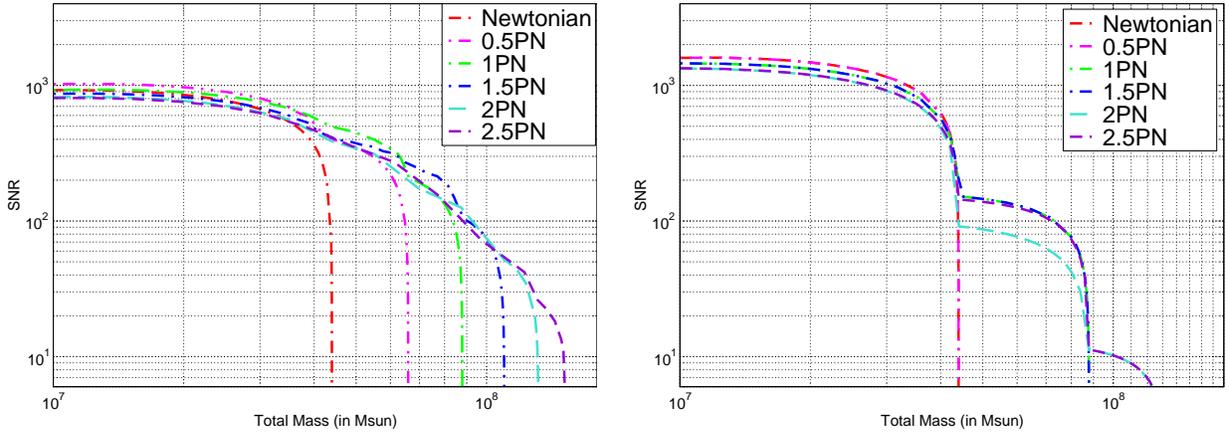

\centering
\includegraphics[width=3.1in]{SNR-1.eps}
\hskip 0.35 true cm
\includegraphics[width=3.1 in]{eqmassSNR.eps}
\caption{SNR versus total mass for successive PN amplitude-corrected
waveforms and 3.5PN phasing. The left panel corresponds  to a
 mass-ratio of 0.1 while
the right panel plots the same quantities  for mass-ratio
of 1  (equal mass systems). 
The $(2n+2)^{\rm th}$ harmonic first appears at the $n$th PN order. 
For a binary of given total mass, the upper cut-off of
the $k^{\rm th}$-harmonic of the orbital frequency 
in the frequency domain is  proportional to $k$
and inversely proportional to the total mass. 
As the mass increases the upper cut-off for the $2^{\rm nd}$ 
harmonic falls below
 the lower cut-off of the LISA detector, 
leading to a zero value of SNR due to the RWF. The higher harmonics still 
enter the sensitive bandwidth of LISA and 
 higher PN order waveforms produce significant SNR. 
The 2.5PN waveform has the highest mass-reach, being 3.5 times the mass-reach of the RWF. 
In the equal mass case displayed in the right panel,
the differences in harmonic content of different PN order waveforms are
 more pronounced, as odd harmonics are absent. 
Sources are at a luminosity distance of 3 Gpc with fixed angles given by 
$\theta_S=\cos^{-1}(-0.6)$, $\phi_S =1$, $\theta_L=\cos^{-1}(0.2)$, $\phi_L=3$. }
\label{fig:eqUneq}
\end{figure*}

Finally, we would like to note an important point not explicitly
mentioned in Ref.\ \cite{ ChrisAnand06}. As the difference between the polarisation
phases of successive PN corrections of the same harmonic tend to be nearly
$\pi$, alternate PN corrections necessarily interfere constructively.
Hence there are positive contributions also from the interference terms.
Now, the numerical value of the contribution to SNR from each of these
terms depends on the magnitude of the polarisation amplitude and the power
of $(Mf)$. It can be checked that for all allowed values of
$(\nu,\theta,\phi,\psi,\i)$ , the polarisation amplitudes are roughly
of the same order in magnitude. Consider the
Newtonian term of the dominant harmonic and its interference
with the first two corrections at $1$PN and $1.5$PN. The Newtonian
term will be out of phase with the $1$PN term, but in phase
with the $1.5$PN one. The two corresponding interference terms
will contain powers of $(Mf)^{2/3}$ and $(Mf),$ respectively, and since
they have the same frequency span, the absolute numerical value of
the contribution to SNR from the former will be more since $(2\pi M f)^{1/3}$
will always be less than ${1}/{\sqrt6}$.
Numerical values of contributions from interference between
higher PN corrections of the second harmonic successively decrease.
The same argument applies for all the other harmonics, and
thus, inclusion of amplitude corrections will lead to an overall
reduction in SNR.

The first column of Table\ \ref{snrtable} clearly demonstrates the
effect of higher harmonics on SNRs. The increase in SNR for the $0.5$PN 
waveform (with respect the RWF SNR) is also easily explained by noting that 
the $0.5$PN correction 
only adds (apart from harmonic mixtures) two direct terms to 
$|\tilde{h} (f)|^2 $, corresponding to the first and third harmonics
($n=1$, $n=3$). Clearly, from the discussion in the previous subsection, 
the $0.5$PN waveform will have a higher SNR than the RWF, independent
of the binary parameters.

For $10^3 \lsim M \lsim10^5 M_\odot$, the difference between the RWF and 
the FWF is not visible on the scale of Fig. \ref{SNR-mass}
because for this mass range all the direct and interference
terms corresponding to harmonics higher than the dominant ones, 
which are scaled by higher powers of  $(Mf)$, are negligible.
 
\subsection{Visibility of systems with $\mathbf{M>4 \times 10^7 M_\odot}$}

In their analysis of the implications of the FWF for ground-based detectors
Van Den Broeck and Sengupta  \cite {Chris06, ChrisAnand06} pointed out
an interesting effect due to higher harmonics. 
An analogous effect is found in the case of LISA in spite of the 
additional amplitude and Doppler modulations that exist in this case.

Normally, the  harmonic at twice the orbital frequency dominates the SNR.
However, when the dominant harmonic fails to reach the LISA band the higher 
harmonics become important, which transpires for masses greater than  
$ 4\times 10^7 M_\odot$. Even though the second harmonic
falls below  the lower cut-off $f_{\rm s}$ of the LISA bandwidth, 
the $k$th harmonic, $k>2,$ that has power up to a frequency  $k\,F_{\rm LSO}$,
might cross $f_{\rm s}$ and  produce a significant SNR. 
Of course, the $k$th harmonic would fall below the LISA sensitivity band
for masses which satisfy the equality  $f_{\rm s} = k\,F_{\rm LSO}$. 
Thus, higher PN order waveforms, which bring in higher
harmonics, are capable of producing a significant SNR, even when the
RWF fails to produce any. 

Let us examine this in a little more detail starting from the values
of mass where the second harmonic dominates and the RWF is adequate.
Eventually, for  larger  values of the total mass, the inequality  
$f_{\rm s} \ge 3F_{\rm LSO}$ becomes true. Then the $0.5$PN waveform, 
which contains  the first and the third harmonic, terminates before 
reaching  $f_{\rm s} $ and consequently the SNR due to  the $0.5$PN 
waveform goes to zero. SNRs for different PN waveforms for a binary
whose dominant harmonic falls below $f_{\rm s}$ and the third harmonic
has a small span in the LISA bandwidth is given in the second column
of Table \ref{snrtable}. Note that for the $5.5(10^6-10^7)M_{\odot}$ system, 
the $1$PN waveform has a higher SNR than the $0.5$PN one. This is due to the absence 
of the first harmonic and the small span of the third harmonic in the LISA bandwidth. 
Further, the $2.5$PN waveform has a slightly larger SNR compared to $2$PN. This 
is due to the absence of the first and second harmonic and the small contribution from
the third harmonic, all of which contribute interference terms due to their $2.5$PN
corrections. However, this increase is marginal, and is not generic. We have 
explicitly checked by choosing different angles that there can be a small 
decrease also. 
The detailed  results for LISA  are summarised in   Fig~\ref{fig:eqUneq}. 
We see that for masses for which the $1$PN waveform fails to reach the LISA 
bandwidth, the higher PN order amplitudes are capable of producing SNRs as 
high as 100!  Thus, the use of the FWF will enable LISA to make
observations of SMBHs in the astrophysically interesting mass-regime, which
would not be possible had one used only the standard RWF.

Using the expression for $F_{\rm LSO},$ it is simple to argue that the mass 
reach for the 2.5PN FWF, which has the seventh  harmonic of the orbital frequency,  
is $7/2$ times the RWF (around $1.5\times 10^8 M_\odot$).
The above ratio, of course, depends on the assumption that the
Schwarzschild (test particle case) LSO frequency will not be
very different from the LSO frequency in the comparable mass case.
 
We conclude with a discussion of a minor, but clear, feature
seen in Fig.~\ref{SNR-mass} for LISA, but not present for the ground-based
detectors, concerning the relative values of the SNR obtained using 
the RWF and the FWF. For most of the mass range probed the RWF
overestimates the SNR relative to the FWF; however, the figure
clearly shows an anomaly  for masses around $\sim 2 \times 10^6 M_\odot$. 
To understand this, we first note that the dips in the two curves 
in Fig~\ref{SNR-mass}, are due to the bump in the LISA noise-curve \cite{BBW05a}
just above $10^{-3}$ Hz. This
bump is due  to the domination of white-dwarf confusion noise over
instrumental noise  and lies just below the most sensitive frequency region
($\sim 3\times 10^{-3}$Hz - $ 2\times 10^{-2}$ Hz) of the LISA band. Below $3
\times 10^{-3}$ Hz, the noise increases sharply till one reaches 
the bump. For binaries of mass greater than $1.5\times 10^6 M_\odot$, the
frequency span of the dominant
harmonic ends just around the bump and the sensitive region of the LISA band
is beyond the span of this harmonic. 
However, higher harmonics incorporated in the FWF
are able to reach the sensitive part of the noise curve. 
This leads to higher SNR for the FWF relative to the RWF.
This reversal of trend continues up to masses  $4\times 10^4 M_\odot$. 
Above this mass, the frequency span of the seventh harmonic
ends before the sensitive region of the LISA band and the general trend
is restored.

           For still higher mass values, the SNRs due to the
RWF and the FWF both increase until the second harmonic fails to reach the LISA
band. This is due to the overall scaling of the waveform with the total
mass. At such high values, it is able to compensate both  for the decreasing
frequency span and  the higher noise of the detector in this  frequency range.

\begin{figure*}[t]
\centering
\hskip 0.35 true cm
\includegraphics[width=3.1in]{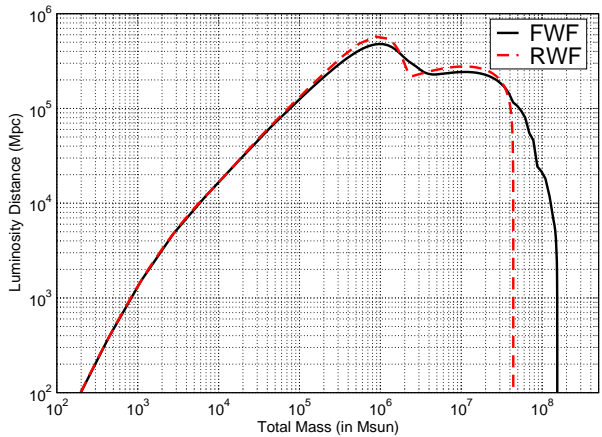}
\caption{ Luminosity Distance (in Mpc) versus total mass for a fixed 
SNR of 10. The systems have mass-ratio of 0.1. 
The distance reach can be as large as 500 Gpc  for
systems where the second harmonic 
enters the LISA bandwidth. 
Systems undetectable by the RWF (of mass around $10^8M_\odot$) 
can be detected by the FWF at distances up to 10 Gpc.
 The location and orientation of the sources are the same as in the earlier figures.}
\label{distance-reach}
\end{figure*}

\subsection{Effect of higher harmonics in the equal mass case}    
In contrast to asymmetric systems discussed so far,
for systems of equal mass {\it all} odd harmonics are absent.  
Consequently, for symmetric systems the
mass-reach of the 2.5PN FWF will be only $3$ times the mass-reach of the RWF.
Further, from the right panel of Fig.~\ref{fig:eqUneq}, 
it is clear that the $0.5$PN and the $0$PN, or RWF, are identical, as are
the $1$PN and $1.5$PN waveforms. Thus the decrease in SNR for
the higher PN order waveforms with increasing total mass is  more pronounced
than in the unequal-mass case. We also note that for masses for which the second
harmonic fails to reach the detector bandwidth, the $2$PN waveform has a lower SNR
than the $2.5$PN waveform. This can be explained by noting that for these masses only
the fourth and sixth harmonics enter the LISA bandwidth. The $2$PN waveform contains
the leading term of the fourth harmonic at $1$PN and its $2$PN correction, which interfere
destructively. However, inclusion of the $2.5$PN amplitude correction leads to a
constructive interference term between the $2.5$PN correction and the $1$PN term
which is responsible for increasing the SNR for the $2.5$PN waveform.

It is interesting to note that the computation of the 3PN GW  polarization which
will introduce an harmonic at  $8\Psi$ will be quantitatively  more
significant  for the equal mass case as the mass reach will be better
by $33\%$ relative to the 2.5PN FWF as opposed to the unequal mass case where it is 
only $14\%$! This provides one motivation for work in progress 
towards  the  computation of the  3PN accurate GW polarizations~\cite{BFIS07}.

\subsection{Variation with mass ratio} 
Since the mass reach depends only on the total mass, the trends
remain the same for different values of mass ratios. Fig.~\ref{SNR-mass}
compares the variation of SNRs with mass for mass ratios of $0.1$ and
$0.01$. If the SNR is dominated by the second harmonic,
the SNR is  smaller for more asymmetric systems  by an  overall
factor of $\nu,$ where $\nu=m_1\,m_2/m^2$.
However, once the second harmonic fails to reach the sensitive  bandwidth
of LISA, the more asymmetric systems have a dominant contribution from the odd
harmonics which scale  by a further factor of $\sqrt{1-4\nu}$, 
which is larger for more asymmetric systems. Thus the decrease in SNR
for the FWF with an  increase in  the total mass is
less steep for more asymmetric systems.

\subsection{Distance reach with  the 2.5PN  FWF}
Next, we compare the distance-reach of the RWF and the 2.5PN  FWF. The
results are shown graphically in Fig.~\ref{distance-reach} and are similar in
appearance to the mass-reach plot. The mass-reach of the RWF is $\simeq
4\times 10^7 M_\odot$. For a system of total mass  $5\times 10^7
M_\odot$, the plot shows that LISA can detect such binaries
 with an SNR of 10 at a
luminosity distance of 100 Gpc ($z\simeq 15$). SMBHs
 of total mass $\sim 10^8 M_\odot$, 
 not even  observable using RWF templates,
have a distance-reach as high as 10 Gpc ($z\simeq1.5$)
with an  SNR of 10. 

Proposals  to  extend the frequency band-width of LISA up to $10^{-5}$ Hz have been
discussed. In that  case, the FWF can increase the mass-reach of LISA 
to even around $10^{9}M_\odot$. 
More specifically, LISA can then observe a  $10^{9}M_\odot$ system with an 
SNR of about 30 at 3 Gpc, if it uses templates based on the 2.5PN
 FWF for data-analysis.

 \begin{figure*}[t]
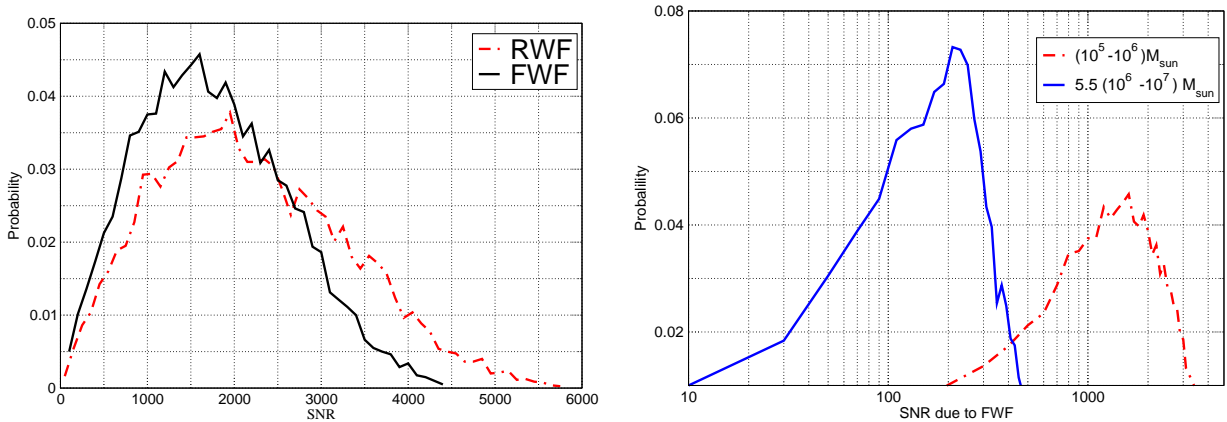

\centering
\includegraphics[width=3.1 in]{anglavg5_6.eps}
\hskip 0.35 true cm
\includegraphics[width=3.1 in]{anglavgecombined.eps}
\caption{Distribution of SNR with sources randomly located and oriented in the sky.
The left figure plots SNRs due to both RWF and FWF for a binary of mass
 ($10^5-10^6) M_\odot$. For this mass, the most probable SNR for the FWF
is lower than the most probable SNR for the RWF,
 like the trend shown in Table~\ref{snrtable}.
 The right figure compares the SNRs due to the FWF 
for binaries of mass ($10^5-10^6) M_\odot$  and $5.5(10^6-10^7) M_\odot$
}
\label{angle-average}
\end{figure*}

\subsection{Sensitivity of SNR to source location and orientation}

 All the   results for SNR using the  amplitude-corrected waveforms quoted
earlier in this paper have been for a fixed choice of location
and orientation of the source [defined by the angles($\theta_S,\,\phi_S,
\,\theta_L,\,\phi_L$)] with respect to the barycentre coordinate 
system.  To conclude our present analysis, in this section we look
into the  variation in  the value of SNR  for sources at
various locations in the sky and various orientations.
To this end, we consider a 
collection of sources randomly oriented in the sky and study the 
probability distribution of their SNRs. 
The results of our simulations (consisting of 8000 random realisations
of the angles involved) 
are shown in Fig.~\ref{angle-average}. 
From the left panel of Fig.~\ref{angle-average}
we see that the most probable SNR due to the FWF for a  $(10^5,\, 10^6) M_\odot$ binary 
is less than the most probable SNR due to the RWF, indicating that this trend
is independent of the source location and orientation. In the right panel we
see that  a binary of mass  $2\times (10^6,\, 10^7) M_\odot$, which is undetectable 
by the RWF, can be observed by the FWF with a most-probable SNR of around $220$.

\section{Summary}\label{Summary}
The implications of amplitude corrected 2.5PN {\em full} waveforms (FWF) 
for the construction of detection templates for  
LISA are investigated in detail. With the FWF, LISA can observe
sources which are favoured by astronomical observations, but not
observable with restricted waveforms (RWF). This includes binaries 
in the mass range $10^{8}-10^9M_\odot,$
depending on whether the lower cut-off for LISA is chosen
to be at $10^{-4}$ Hz or $10^{-5}$ Hz. With an SNR of 10, these
systems can be observed up to a redshift of about 1.5.
The computation of the 3PN polarization, which will introduce an
harmonic at $8\Psi$ (i.e. four times the dominant harmonic),
in addition to  the existing harmonics, could enhance the mass reach 
for  equal mass binaries by $33\%$ and unequal mass binaries by $14.3\%$.

The implication of the FWF for parameter estimation will be far more
important than the extension of LISA's mass-reach reported here. From the
work of Van Den Broeck and Sengupta in the context of ground-based
detectors \cite{ChrisAnand06b} it is already clear that most parameters will
be estimated with errors $\sim$ ten times smaller as compared to RWF. This raises
the interesting possibility that binary SMBH coalescences might be located on
the sky with accuracies good enough for optical observations to
identify the galaxy cluster and measure its red-shift.  Needless to
say that this improved estimation of source properties
will have important consequences in shedding light on the dark energy,
better understanding of SMBH formation and evolution, structure formation,
etc., and is currently under investigation. 

In this work we have confined ourselves to only non-spinning black-holes 
ignoring the effect of spin-orbit coupling at 1.5PN~\cite{KWWi93} and
2.5PN~\cite{BBuF06} and spin-spin effect at 2PN order~\cite{PW95}. 
The effect of spin is expected to be astrophysically significant and it
 is important to revisit the present analysis including spin in the future.
Though partial results for GW polarisations including spin do exist, 
a more exhaustive exercise would be necessary before the FWF required for 
this work is available. The problem will also be
more complicated due to modulations arising from spin-orbit and spin-spin 
couplings which would need to be addressed.

In this work we also restricted to the inspiral phase and used a physical 
picture of the LSO that is based on the test-particle limit.
For comparable masses, the notion of LSO is not as sharp, or unique,
and hence our results are probably idealized limits of the
real situation. Numerical relativity \cite{Bruegmann03,Pretorius05, BCCKM06} is 
maturing over the past couple of years and could soon provide waveforms 
for late inspiral and merger. It should then be possible to compare the
results of such numerical templates with those studied in this paper
to provide a better understanding of how higher harmonics facilitate
the mass reach of our detectors.

\begin{acknowledgments}
KGA  acknowledges the Cardiff university for hospitality  during the initial
stages of this work and thanks  Chris Van Den Broeck and Anand Sengupta
for useful discussions on data-analysis with the FWF for ground-based detectors. 
KGA also acknowledges VESF. BRI, BSS and   SS thank
 the Institut Henri Poincar\'e  and BRI the Institut des Hautes Etudes
 Scientifiques  for hospitality during the initial stages of this work.
All the calculations reported in this paper are performed with
{\it Mathematica}.
\end{acknowledgments}

\bibliography{./ref-list}
\end{document}